# ttH, ttZ and ttW measurements at LHC and top quark compositeness

François Richard

Laboratoire de l'Accélérateur Linéaire, IN2P2–CNRS et Université de Paris–Sud,
Bât. 200, BP 34, F–91898 Orsay Cedex, France.

***Abstract***: This note is an attempt to interpret some excesses – not yet significant due to systematics – observed by ATLAS and CMS in various analyses related to the standard channels ttH, ttZ and ttW. It is argued, within a composite interpretation of top particles, that such excesses are not necessarily related to these channels themselves –although this is not excluded – but due to the underlying presence of either vector-like heavy quarks t',b' or to $t\bar{t}t\bar{t}$ final states as predicted in composite theories. The outcome of this discussion is that although it will not be easy to reach an exclusive interpretation, the data collected at 13 TeV may establish the origin of this effect as coming from the $t\bar{t}t\bar{t}$ topology.

## I.   Introduction

So far no evidence for new physics has emerged from ATLAS and CMS, neither from standard channels measurements nor from searches. In this article, it is argued that the ttH, ttW and ttZ measurements can serve as a sensitive test for the presence or absence of BSM physics from several composite models considered in the literature. The reason is that the top quark, the Higgs bosons and its avatars, the longitudinal components of W and Z, could carry a great deal of compositeness. While it will take an e+e- collider, with its accurate measurements of Higgs and top couplings [1], [2] and [3], to bring the ultimate proof of the composite character of these objects, LHC could soon observe direct signals related to composite sources.

On the experimental side ttH and ttW show, at 8 TeV, an excess both in ATLAS and CMS.

Given that these measurements are not exclusive, meaning that it is impractical to fully reconstruct the content of a given event, one cannot unambiguously understand the source of these effects. In the case of ttH, one can speculate that the Yukawa coupling is in excess to the SM value due, for instance, to a mixing between the top quark and a new heavy quark t', as was done in [4]. One can alternatively speculate on the presence of BSM final states which pass the loose selections used for this channel.

While a 4$^{th}$ generation of SM quarks and leptons is excluded by a combination of LEP/SLC and LHC precision measurements [5], nothing prevents the occurrence of heavy vector-like fermions. In fact they naturally appear, for instance, in Randal Sundrum (RS) [6] and Little Higgs [7] models.

If these heavy quarks are present in the LHC data, they could contaminate the corresponding ttH topologies under consideration. For instance, if one produces a pair of b' with b'->tW, these events could provide 4W final states giving like-sign leptons which are used to select ttH and ttW. At present no direct evidence for these heavy quarks has been provided by ATLAS nor CMS while there are a few yet modest excesses which could be interpreted by their contribution.

Another way to feed these ttX channels would be to produce four top final states $t\bar{t}t\bar{t}$. While the SM contribution to this process is very small, at the fb level, one predicts coloured resonances, pair produced and decaying into top pairs as in [8] or [9]. Alternatively there could be an enhancement of the small SM cross section for the production of $t\bar{t}t\bar{t}$ due to final state interaction as expected in composite models [10]. This topology produces 4W final states feeding the like-sign lepton signals and 4 b jets searches.

ATLAS has released a ttH analysis at 13 TeV with 13 fb-1 integrated luminosity which confirms the excess observed at 8 TeV [11].

It is the purpose of this note to discuss these aspects, mainly on the basis of the ttH, ttZ and ttW ongoing measurements for which there are precise SM predictions, although these measurements suffer from large uncertainties coming from background estimates.

## II. Possible sources contaminating ttH, ttZ and ttW

### II.1 Heavy quarks

Their production cross sections are fairly universal for pair production which go through the same mechanism as top pair production. For masses mQ~800 GeV which, roughly speaking, correspond to the present reach of LHC, one expects, at 13 TeV, an increase of the gluon-gluon luminosity by more than an order of magnitude with respect to 8 TeV, reaching a cross section above 200 fb (for mQ=900 GeV this cross section falls below 100 fb). The ttH channel, as we shall discuss shortly, is populated by similar topologies and has a similar cross section, meaning that both channels need to be considered to reach an interpretation of an excess. This is also the case for ttW and ttZ.

Which are the decay modes of these heavy quarks? [12] gives a detailed description of this phenomenology. These heavy quarks can be either singlets, doublets of triplets. In the following, the discussion will focus on quarks carrying the same quantum numbers as t and b and they will be called t' and b' but there is of course the interesting possibility to produce heavy quarks with exotic charges 5/3 and -4/3.

These heavy quarks receive their masses from a common mechanism and tend to be degenerate in mass if one can neglect the electroweak contribution, as is the case in the mass range under consideration which is above 600 GeV. They are therefore likely to be simultaneously present. This also means that they do not cascade into each other. In most models they will decay into the t or b, another reason to called them t' and b'.

These branching ratios into charged current modes like b'->tW and neutral current modes like t'->tH and t'->tZ depend on unknown mixing angles, in particular neutral currents can be completely suppressed or dominant. If they are suppressed, a heavy b' will decay predominantly into tW feeding topologies where excesses have been observed as discussed in the next sections. If, on the contrary, neutral current dominate, on expects that t'->tH and t'->tZ, with equal branching ratio (equivalence theorem), which will feed channels where no excess is observed.

These heavy quarks, if present, could play a major role in the Higgs sector since, for instance, they could contribute, through loops, to the two gluon coupling. For what concerns Z couplings, measured at LEP1/SLC, a b' could mix with b and influence the Zbb width. These aspects, not dealt here, are discussed in [12].

This picture can be modified in two mays. There could be an increase of the t't' cross section, in the presence of a gluon KK recurrence provided that its mass is of order 2 TeV, which is still not excluded by present searches. One can also produce t't final states, provided that t' mixes significantly with t. If t'->tH, the resulting final state would be identical to ttH.

## II.2 Four top final states

For a general presentation and references, see [13]. Only a few specific points will be recalled.

### II.2.1 Kaluza Klein resonance

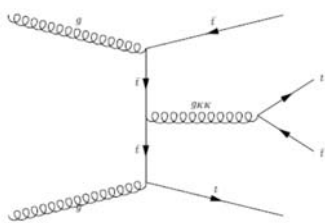

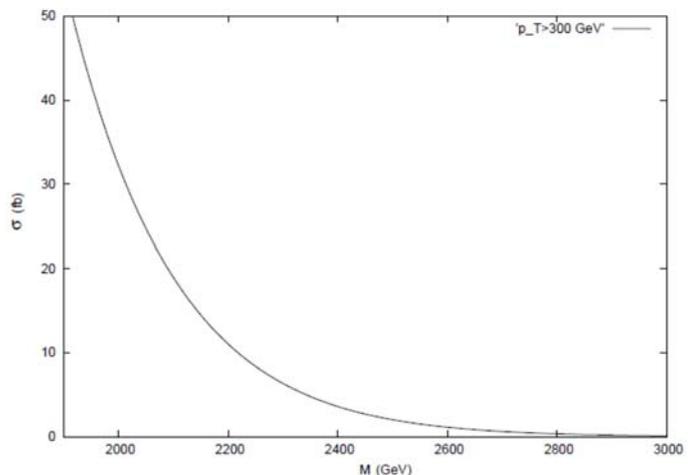

Figure 1: Predicted cross section for the production of a KK resonance versus its mass for LHC at 14 TeV. A transverse momentum cut is applied on the top quarks.

$t\bar{t}t\bar{t}$ production can appear in RS through the production of a KK gluon [8]. At 13 TeV and for a KK resonance with a mass of ~2 TeV the cross section reaches 50 fb assuming a branching ration into top pairs close to 100%.

## II.2.2 Composite tops

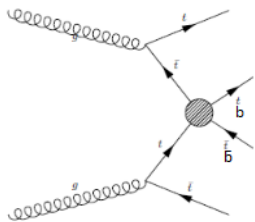
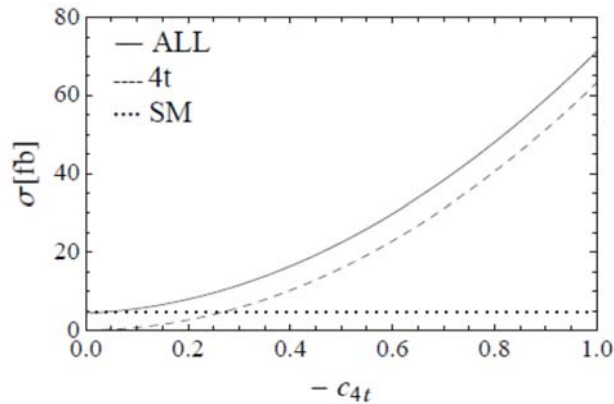

*Figure 2: Predicted cross section for the 4 top production at 14 TeV in a composite model (full line) versus an effective coupling parameter. The dotted line corresponds to the SM contribution.*

Composite tops can re-interact in the final state, enhancing the process pp->$t\bar{t}t\bar{t}$ as schematically shown in above diagram. This topology has a specific feature which may help in its identification: there are two interacting tops and two spectator quarks, these being forward emitted. One can also have pp->$t\bar{t}b\bar{b}$ [10]. The expected cross section at 13 TeV can reach ~50 fb.

## II.2.3 Colored resonances

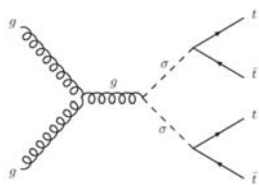

They appear, within SUSY, as scalar gluons but also, within composite models, as in vector-like confining theories or extra-dimensions models [13]. As coloured resonances, they strongly couple to gluons as shown by the diagram. The cross section depends on the mass of these resonances. In [9] one predicts a cross section at the 100 fb level for TeV resonances. This does not include the branching ratio into top pairs which is at the 10% level, hence a $t\bar{t}t\bar{t}$ cross section of ~1fb.

## III. The ttH channel

At 8 TeV, the ttH SM cross section is ~130 fb. An excess was observed in this channel as recalled in figure 3 [14]. This excess is more marked in CMS than in ATLAS although compatible. When combined the two experiments reach a statistically significant excess but, adding systematics, one falls below 2 s.d. This means that the future of this analysis requires a better understanding of systematics. Figure 3 also indicates that the ratio bb/ZZ is below the SM, implying therefore that the detection using bb decays from H (so called topology 3 in what follows) is not favoured.

At 13 TeV the ttH cross section is ~200 fb. An excess is also observed as shown in figure 4 from [13] but again systematics prevent a conclusive statement. This search uses 3 topologies defined below:

- In the first topology, H is selected through its decay into **two photons** plus a lepton and two jets, one of them being a b jet. A small deficit is observed.
- In the second topology, related to H->WW, one selects **two like-sign leptons** with or without a tau decaying hadronically. This topology has the smallest background. Both channels show an excess.
- In the third topology, one requires at least **3 b jets** meaning that H has to decay into two b. An excess is observed with two lepton of opposite signs.

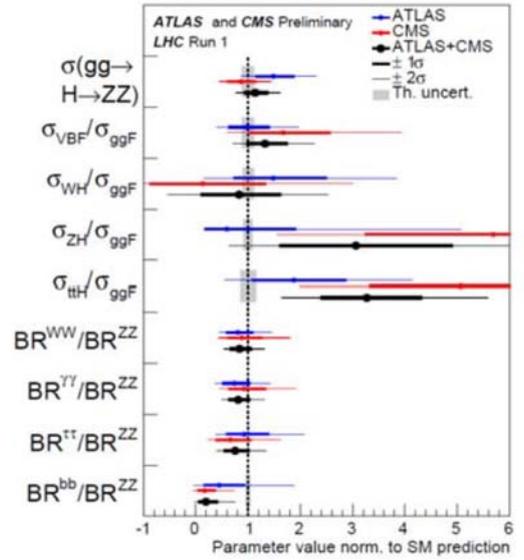

Figure 3: Cross sections and ratios of branching ratios for the Higgs boson from ATLAS and CMS and combined results at 8 TeV. The vertical line corresponds to the SM prediction.

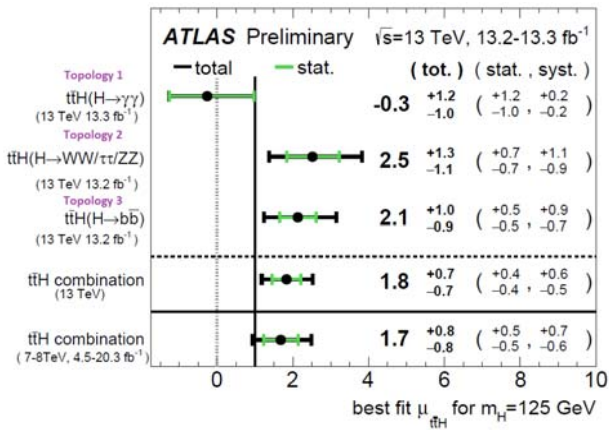

Figure 4: Signal strength measurements for the three topologies and their combination compared to expectations (vertical black line) at 13 TeV. Small errors are statistical only. The 8 TeV combination is recalled.

One sees that, with the exception of the first topology, no explicit selection insures the presence of H. This even partially applies to the first topology since this signal is not free of combinatorial background.

The small deficit in the first topology seems to speak against an important contribution of decays like t'->tH or b'->bH. It also speaks against an increase of the Yukawa coupling Yt which was advocated in [4] to explain the global excess in ttH. Again, these statements are to be taken with a grain of salt at the present level of statistical significance.

For the second topology, the large branching ratio of H into WW, ~20%, provides 4W final states which allows to use the powerful like-sign leptons signature with rare appearance in the SM backgrounds. b'->tW and $t\bar{t}t\bar{t}$ provide 4W final states which give like-sign lepton for this topology.

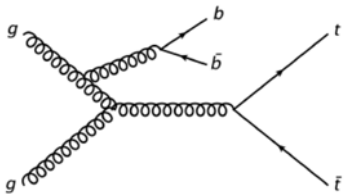

The third topology is heavily contaminated by the QCD effect shown in the diagram. Demanding two additional opposite sign leptons helps in eliminating various QCD contributions but the top pair production itself gives a huge background not easy to estimate and responsible for systematics.

Heavy quarks with t'->tH and H->2b or $t\bar{t}t\bar{t}$ can fulfil the selection criteria for this topology.

## IV. A search for four top states

In this section one discusses a search for such states, of interest to understand a possible contamination for the ttX final states.

The histogram shown in figure 5 is from [15]. It is obtained with 8 TeV data by requesting:

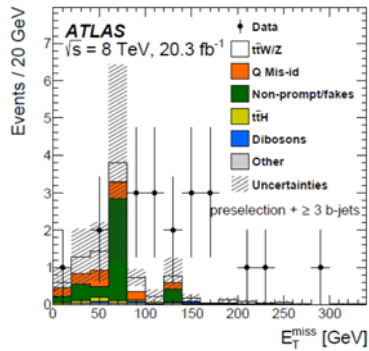

- Two like-sign leptons
- At least 3 b jets

Figure 5 shows the transverse missing energy $E_T$miss distribution. This topology can easily be satisfied by $t\bar{t}t\bar{t}$.

If one selects $E_T$miss>100 GeV, keeping about 2/3 of the $t\bar{t}t\bar{t}$ signal as deduced from [13], the number of predicted background events, about **2**, is clearly not compatible with the **14 observed events**. Depending to which statistical school you belong – Bayesian or frequentist – the claim can be between 3.5 and 5 standard deviations, which appears quite significant. In any case the data collected at 13 TeV will tell.

*Figure 5: Missing transverse energy distribution for events with at least 3 b jets and two like sign leptons as seen in ATLAS at 8 TeV*

Note that, aside from the fake events, the SM background at large $E_T$miss comes from the ttW topology (which can only satisfy the demand for at least 3 b jets through b tagging misidentification).

The excess of events, in the $t\bar{t}t\bar{t}$ interpretation, corresponds to a measured cross section of about 19 fb, assuming a reconstruction efficiency estimated at the ~6% level, which includes the leptonic branching ratios and the efficiencies for the b tagging and lepton selections. From [13] one can estimate that about 2/3 of the four top $t\bar{t}t\bar{t}$ events pass $E_T$miss>100 GeV and deduce σ$t\bar{t}t\bar{t}$ **=29±8 fb** where the error is statistical.

The same reference sets an upper limit on this cross section of 70 fb, in excess to the expected limit of 20 fb. In [16], no excess is observed, but this search does not consider same sign leptons associated to b jets. It sets an upper limit on $t\bar{t}t\bar{t}$ at 23 fb, therefore without tension with the value deduced from figure 5. The same is true for the upper limit at 32 fb from [17].

Such a large cross section enables the $t\bar{t}t\bar{t}$ process to contaminate the ttH channels, with the exception of the first topology.

The theoretical predictions for the $t\bar{t}t\bar{t}$ cross section presented in section II.2 which are valid at 14 TeV, tend to fall below this value but there is room for large uncertainties in these models.

The ttWW contribution, coming from b'b' events, only provides 2 b jets, while this selection asks for at least 3 b jets. It is however possible to have fake b jets, therefore this hypothesis cannot be completely ruled out.

Again, at 13 TeV, there should be no problem to confirm or discard the excess shown in figure 5, given that the $t\bar{t}t\bar{t}$ cross section will increase by more than 2 at this energy.

## V. The ttW channel

At 8 TeV, the SM cross section is 200 fb. A small excess is observed in ttW by both experiments, [18] and [19], as shown in figure 6. This excess is seen with same sign leptons and could come from b'b' or $t\bar{t}t\bar{t}$ final states.

The ttZ mode is close to the SM prediction, which does not favour large branching ratio for t'->tZ.

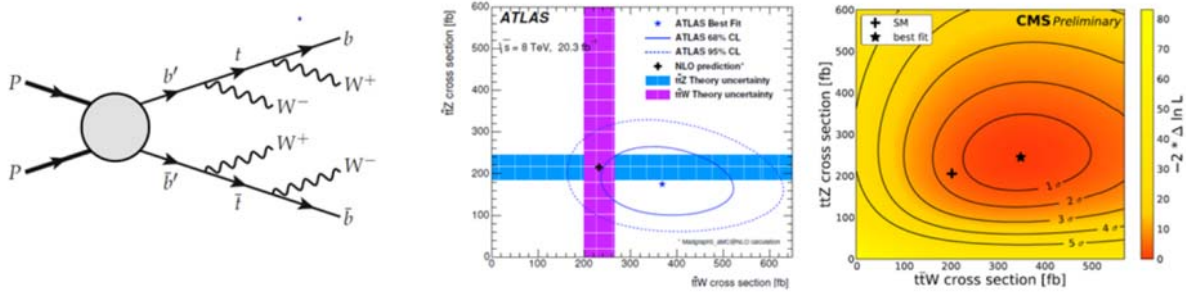

*Fig 6: ttZ and ttW results for ATLAS and CMS at 8 TeV*

At 13 GeV one expects a cross section of 0.84 pb for ttZ and 0.60 pb for ttW. Based on a small sample, [20] observes 0.9±0.3 pb for ttZ and 1.4±0.8 pb for ttW, showing a similar trend.

## VI. Interpretations

The clearest excesses observed in the ttX topologies, come from same sign leptons, which favours 4W final states either from b'b' with b'->tW decays or from a $t\bar{t}t\bar{t}$ final state. The latter has the advantage to generate 4 b jets, which can also feed topology 3 in the ttH mode and explain the significant excess observed in [15] by combining the same sign leptons and multi b jet requirements. Since both the b'b' and t't' should simultaneously be present, one should also take into account the t' contributions. While the decay t'->bW provides the same signatures as ordinary top quarks and seems therefore irrelevant, t'->tH/tZ would contribute to excesses which are not observed in ttH and ttZ. All this being said with a grain of salt, given the small statistics involved.

If one assumes σ$t\bar{t}t\bar{t}$ =29±8 fb at 8 TeV and given that, in the, SM BR(H->WW)~20% and BR(H->bb)~60%, one predicts:

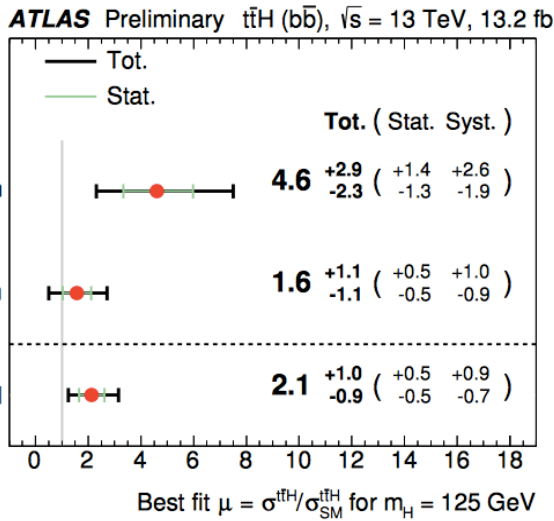

*Figure 7: Results from ATLAS at 13 TeV where the dilepton line corresponds to like-sign leptons.*

μi=σ$t\bar{t}t\bar{t}$ BRi(4t) /σttH*BRi(ttH)+1

where BRi are the branching ratios for a given topology.

This gives **μ2=2.1±0.3** for the second topology and **μ3=1.6±0.2** for the third one, to be compared to μ=1.9±0.8 in ATLAS and μ=2.9±1 in CMS after averaging on the 3 topologies.

For ttW, one predicts **μ=1.4±0.1** instead of the measured value μ=2±0.7 obtained by averaging ATLAS and CMS at 8 TeV.

These agreements could of course be a simple coincidence.

At 13 TeV, the data from [13] show the same average discrepancy as for 8 TeV (figure 4), which can be interpreted by saying that the $t\bar{t}t\bar{t}$ cross section has not increased faster that ttH, with a large margin of uncertainty given the errors. Figure 7 confirms that the predominant excess is observed for like-sign leptons. Dominant syst. errors unfortunately prevent a clear conclusion.

## VII. Summary

This discussion about LHC results on ttH, ttZ and ttW, with a tentative interpretation of a few excesses, is not meant to be conclusive but intends to give a framework in case those indications get strengthened. The data provided so far select a few topologies which rely on inclusive features like leptons, missing transverse energy, b-tagged jets, by no means allowing a direct connection to the ttH and ttW channels. Our purpose was to draw attention on this aspect in order to prevent naïve interpretations of these excesses.

Given the large syst. uncertainties on these measurements, it is not certain that the ttH and ttW anomalies will soon become convincing. More promising appears the type of search described in reference [15] with tight selections relying on a $t\bar{t}t\bar{t}$ final state. Interpreting this excess as the main source for the discrepancies observed in ttH and ttW works beautifully well but this could be a mere coincidence given the very large uncertainties. The alternate model assuming heavy quark production is not ruled out.

From the data collected at 13 TeV and an expected increase of the $t\bar{t}t\bar{t}$ cross section which should be larger than 2, the effect observed in [15] can be verified and provide an important input to the analysis of the ttX channels and, if confirmed, brings a major input to our understanding of top compositeness.

### Acknowledgements
Useful discussions with Gregory Moreau, Roman Poeschl and Daniel Treille are gratefully acknowledged. Special thanks to Andrei Angelescu for reading this note and providing important suggestions.